\documentclass[aps,twocolumn,showpacs,preprintnumbers,amsmath,amssymb,superscriptaddress]{revtex4-1}

\usepackage{hyperref}

\usepackage{amssymb}
\usepackage{color,dcolumn}

\def \Prague{Charles University, Faculty of Mathematics and Physics, Institute of Physics, Ke Karlovu 5, CZ-121 16, Prague 2, Czech Republic}

\usepackage{epsfig}
\usepackage{amsmath}

\begin{document}
\title{Spectral Current Density and Responsivity Scaling for Fourier Transform Photocurrent Spectroscopy}

\author{J. \surname{Kunc$^*$}} 
\author{B. \surname{Morzhuk}}
\author{M. \surname{Shestopalov}} 
\author{T. \surname{Fridri\v{s}ek}} 
\author{V. \surname{D\v{e}di\v{c}}} 
\affiliation{\Prague}
\email{kunc@karlov.mff.cuni.cz}

\date{\today}

\begin{abstract}
We propose and experimentally verify two methods to scale arbitrary units to photocurrent spectral density (A/eV) in Fourier Transform Photocurrent (FTPC) spectroscopy. We also propose the FTPC scaling to responsivity (A/W), provided a narrow-band optical power measurement is available. The constant background of the interferogram provides a precise determination of the photocurrent spectral density. The second method relies on the scaled amplitude of the interferogram. Although the latter method leads to more significant errors, it still provides good order of magnitude estimates of the total photocurrent. We demonstrate the technique on a calibrated InGaAs diode and weak responsivity SiC interdigital sensors. We identify a series of impurity-band and interband transitions in the SiC sensors.
\end{abstract}


\maketitle

\section{Introduction}
The Fourier Transform Photocurrent (FTPC) spectroscopy is a sensitive experimental technique to study semiconductors~\cite{JongbloetsPRB1979,ChenJAP2000,VanecekAPL2002}, hydrogen-related shallow impurities~\cite{LavrovPRB2005,FrenzelPRB2007}, micro-crystalline silicon for solar cells~\cite{VanecekAPL2002}, quantum dots~\cite{PetterssonMicJ2005,HoglundAPL2006}, nano-diamond thin layers~\cite{KravetsDRM2006} or emerging materials such as perovskites~\cite{HolovskyJPCL2017,RidzonovaJMCA2022}. The FTPC is closely related to Photo-thermal Ionisation Spectroscopy (PTIS), which aims to study impurities in high-purity semiconductors~\cite{JongbloetsPRB1979,HikavyyPSSA2005}, including their local field interactions by applying magnetic fields~\cite{ChenJAP2000}. Later, the technique started to be used to measure spectral-resolved photoconductivity~\cite{VanecekAPL2002,PuspitosariRSI2017} with the advantages of higher spectral resolution, higher sensitivity, and faster measurement compared to monochromator-based techniques. However, despite their advantages, the FTPC and PTIS lack a photocurrent and responsivity scaling method.

There were attempts to scale the spectral current density of solar cells \cite{PuspitosariRSI2017}. However, here the authors measured a short-circuit current of the solar cell. That is inapplicable to samples with weak responsivity. Other attempts matched the high-energy spectrum with the transmittance and reflectance data, scaling the FTPC to the absorption units ($\mathrm{cm^{-1}}$), or matching the FTPC and monochromator-based photoconductivity \cite{VandewalTSF2008}.

However, these scaling techniques require high photocurrents or supplementary experiments. The additional experiments diminish the advantages of FTPC spectroscopy, such as fast acquisition times and high dynamical range. These issues typically lead to arbitrary unit scaling of the FTPC spectra~\cite{KromkaPSSA2003,HaenenDRM2004,HoglundAPL2006,GoisAPL2006,KravetsDRM2006,HolovskyPSSA2010,EnorbioJJAP2012,GuentherAPL2014}, and the scaling methodology is still missing~\cite{RidzonovaJMCA2022}.

We propose here two methods to scale the photocurrent spectral density. We also show a simple extension of the photocurrent scaling to determine the sample's responsivity. We derive the two methods from the interference of two planar waves. The interference pattern consists of the constant background and the interference term. The absolute value of the constant background relates to the total photocurrent, provided the amplification ratio of the preamplifier is known. The maximal amplitude of the interference term determines the same integral photocurrent. We show that despite the second method requiring an ideal 50:50 beamsplitter, it provides a good order of magnitude estimate of the total photocurrent. 

We demonstrate two scaling procedures on a calibrated photodiode and SiC interdigital sensor. We start with the calibrated InGaAs photodiode (Thorlabs, FGA21-CAL), where the manufacturer provides a National Institute of Standards and Technology (NIST) traceable responsivity spectrum for each diode. The responsivity of the InGaAs diode approaches 1~A/W, allowing the oscilloscope to measure the photovoltage directly. The measurement with an optical bandpass filter provides an accurate optical power measurement. We also show the scaling procedure on our 6H-SiC (II-VI Inc., semi-insulating, vanadium-doped) interdigital sensors. Here, the photoresponse is six orders of magnitude weaker. Therefore, the oscilloscope's sensitivity is insufficient to scale the spectrum with a bandpass filter. We fabricated the interdigital sensors by a two-step electron beam lithography (Raith 150-Two). The interdigital geometry was defined using dry etching of Ti(5~nm)/Au(15~nm). We used Ti(5~nm)/Cu(85~nm)/Au(15~nm) metalization for wire-bonding. The sensor area is 300$\times$300~$\mu$m$^2$. 

\begin{figure}[t]
\centering
\includegraphics[width=8cm]{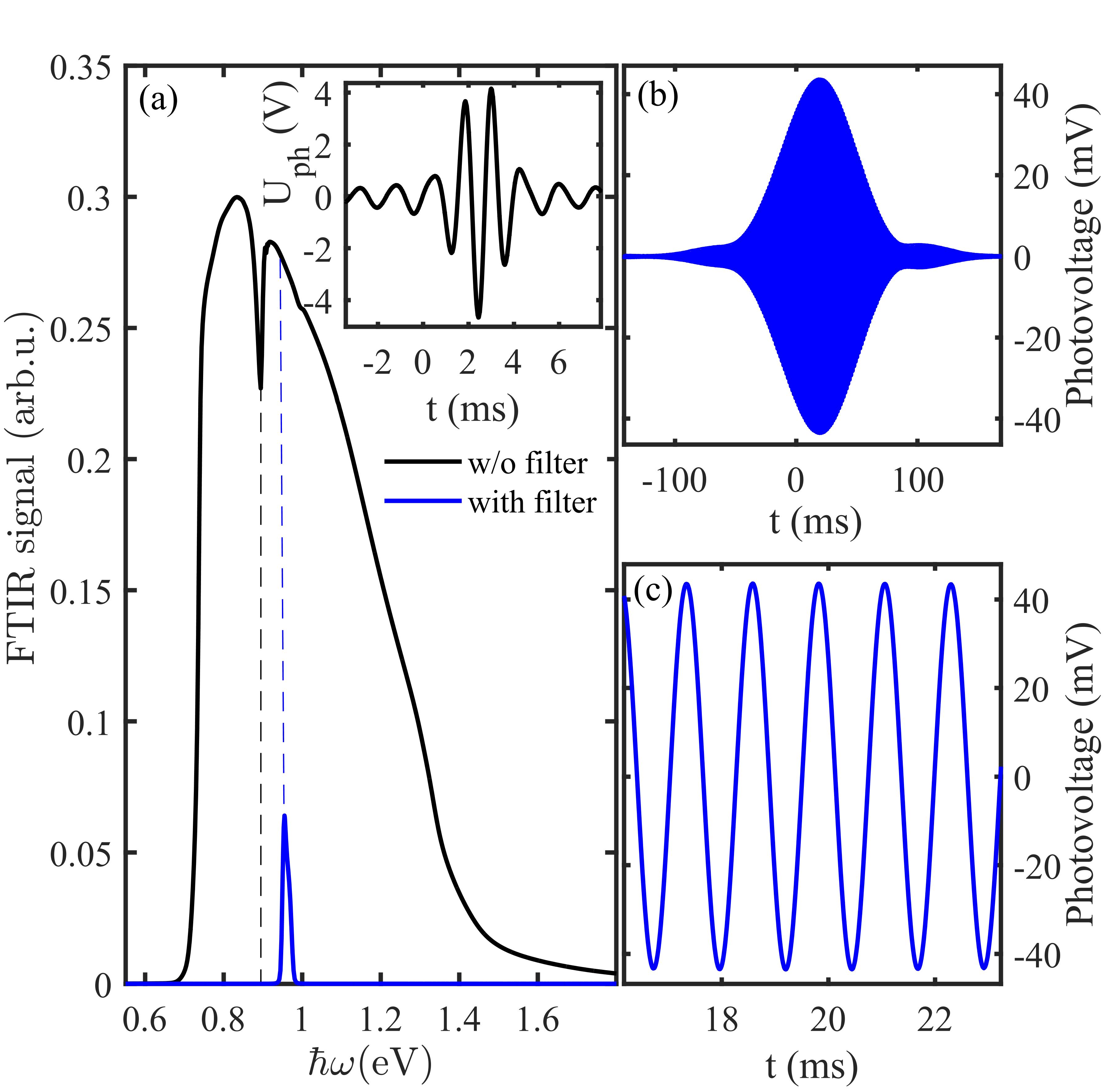}
\caption{ (a) InGaAs FTPS spectrum measured using broadband white light tungsten source (black curve) without and (blue curve) with a bandpass filter (1300$\pm$12)~nm. The inset shows the interferogram without the bandpass filter measured by the oscilloscope. (b) Full interferogram of the InGaAs diode with the bandpass filter. (c) The interferogram detail is near its maximum as measured by the oscilloscope.}
\label{Fig1}
\end{figure} 

The black curve in Fig.~\ref{Fig1}~(a) shows the Fourier Transform Photocurrent Spectrum (FTPS) of the InGaAs diode. The diode is unbiased (photovoltaic detection), and the photocurrent is preamplified by the trans-impedance amplifier (Newport 70710) at the amplification 10$^4$~V/A (bandwidth 8~MHz). We used a tungsten white-light source with an aperture of 1~mm to avoid saturation of the diode and CaF$_2$ beamsplitter. The output photovoltage is carried back in FTIR spectrometer Bruker Vertex80v to perform the Fourier transform. The scanner velocity used throughout our experiments was 1650~Hz. Hence, we are well within the bandwidth of the trans-impedance amplifier at 10$^4$~V/A for all measured wavelengths. The two scaling methods use the relation describing interference of two beams of the intensity $I_1$ and $I_2$, resulting in the interference pattern $I$

\begin{equation}
    I=I_1+I_2+2\sqrt{I_1I_2}\cos(k\Delta z)e^{-t/\tau_\mathrm{coh}}.
    \label{eq:Interference}
\end{equation}

\begin{figure}[t]
\centering
\includegraphics[width=8cm]{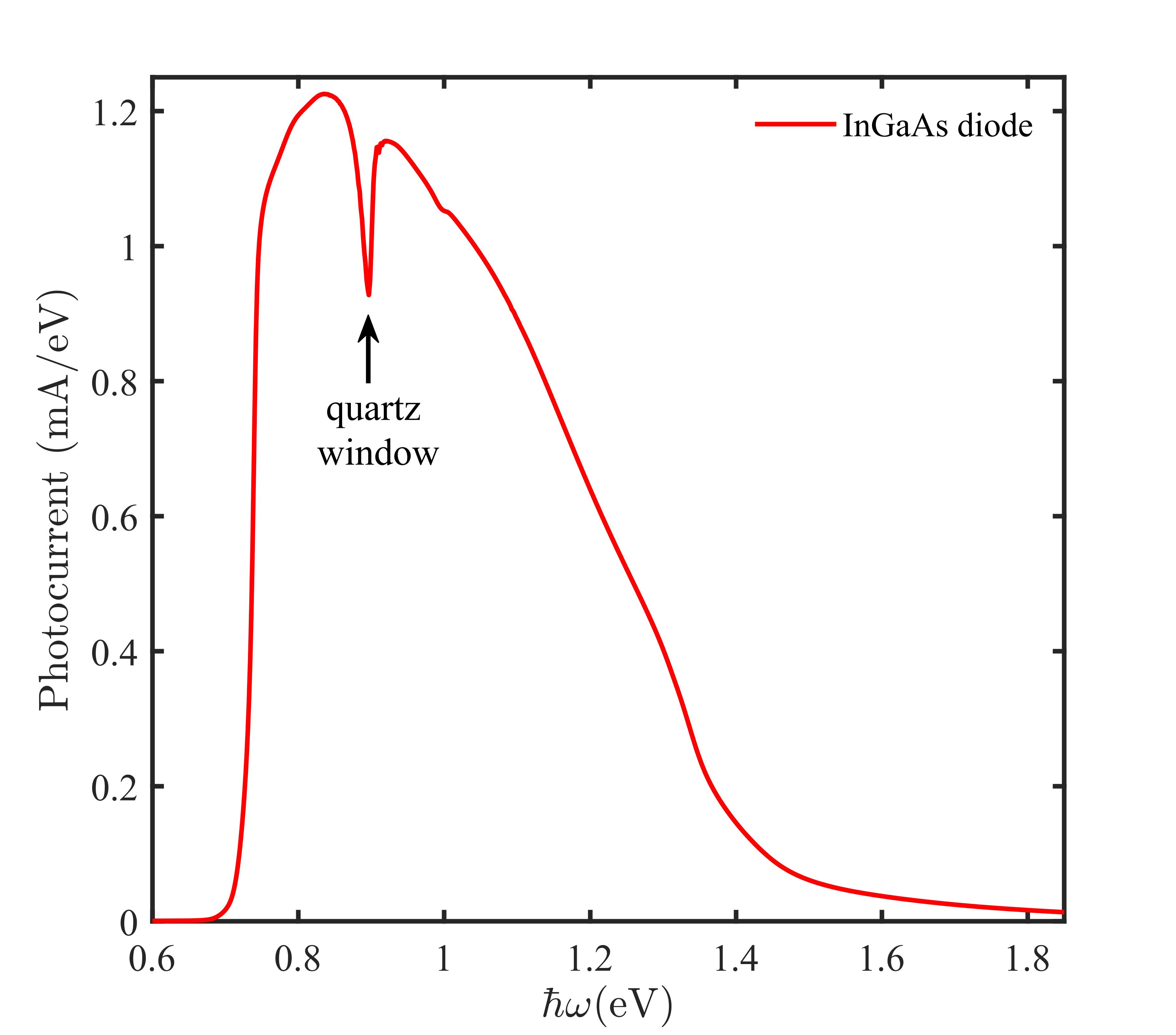}
\caption{InGaAs FTPS spectrum measured using a broadband white light tungsten source scaled to amperes. We scaled the photocurrent by the photovoltage obtained using the bandpass filter (1300$\pm$12)~nm. }
\label{Fig2}
\end{figure}

The wavevector $k=2\pi/\lambda$ describes the spatial modulation of the monochromatic light at the wavelength $\lambda$. The position $z$ of the mirror in the Michelson interferometer leads to intensity $I$ modulation. The coherence time $\tau_\mathrm{coh}$ describes the reduced visibility of the interference with increasing delay time $t$. Equation (\ref{eq:Interference}) provides two methods to calibrate the spectra. First, we measure the constant term $I_1+I_2$ by chopping the light exiting the FTIR spectrometer. We measure this amplified signal by the oscilloscope or by lock-in homodyne detection. The chopped white-light signal from the InGaAs diode (Method 1) leads to the total photovoltage $U_\mathrm{ph, max}^{(1)}$=5780(50)~mV. Second, we measure the prefactor $2\sqrt{I_1I_2}$, assuming $I_1=I_2$. The total photocurrent is then related to the maximum of the interferogram. The inset in Fig.~\ref{Fig1}~(a) shows the interferogram of the preamplified photocurrent. We removed the constant background $I_1+I_2$. The maximal photovoltage reads $U_\mathrm{ph, max}^{(2)}$=4668(5)~mV. The unscaled spectrum $u_i^{\mathrm{exp}}$ is scaled using the measured spectral integrated photocurrent $I_\mathrm{tot}$, giving the spectral photocurrent density $\mathcal{I}$

\begin{equation}
    \mathcal{I}=I_\mathrm{tot}\frac{u^\mathrm{exp}_i}{\sum_{i}u^\mathrm{exp}_idE_i},
\end{equation}

where $dE_i$ is the energy spacing between experimental points $u_i^{\mathrm{exp}}$. The trans-impedance amplification $A_\mathrm{trans}$ (in our case $10^4$~V/A) relates $I_\mathrm{tot}=U_\mathrm{ph,max}^{(i)}/A_\mathrm{trans}$. We show the scaled FTPS spectrum in Fig.~\ref{Fig2}. The quartz windows cause the dip at 0.89-0.90~eV at the output of the FTIR spectrometer. We compare the two scaling methods in Tab.~(\ref{Tab:comparison}). The interferogram maximum underestimates the total photocurrent by 20\%. The underestimation is mostly due to the non-ideal beamsplitter ($I_1\neq I_2$) in the visible spectral range. 

\begin{table*}[t!]
\caption{\label{Tab:comparison} Comparison of the integrated FTIR signal measured with and without bandpass filter for InGaAs calibrated photodiode and our SiC interdigital sensor.}
\begin{ruledtabular}
\begin{tabular}{cccc}
Sample& Total signal & Signal with band-pass filter & ratio\\
\hline
InGaAs - FTIR signal& 6313 arb.u. & 55.9 arb.u. & 113.0 \\
Interferogram $U_\mathrm{max}$  & 4668(5) mV & 43.55(5) mV & 107(3) \\
Interferogram $U_\mathrm{back}$ - pk-pk  &5780(50) mV & 47(2) mV     & 123(5) \\
\hline
SiC - FTIR signal & 48.7 arb.u.    &    0.57 arb. u.  & 85.3 \\ 
Interferogram $U_\mathrm{max}$  &31.1 mV    & NA      & NA\\
Interferogram $U_\mathrm{back}$ - pk-pk &66.2(5) mV & NA      & NA\\
Interferogram $U_\mathrm{back}$ - rms &24.8(5) mV & 0.29(5) mV & 85(15)\\
\end{tabular}
\end{ruledtabular}
\end{table*}

\begin{figure}[t!]
\centering
\includegraphics[width=8cm]{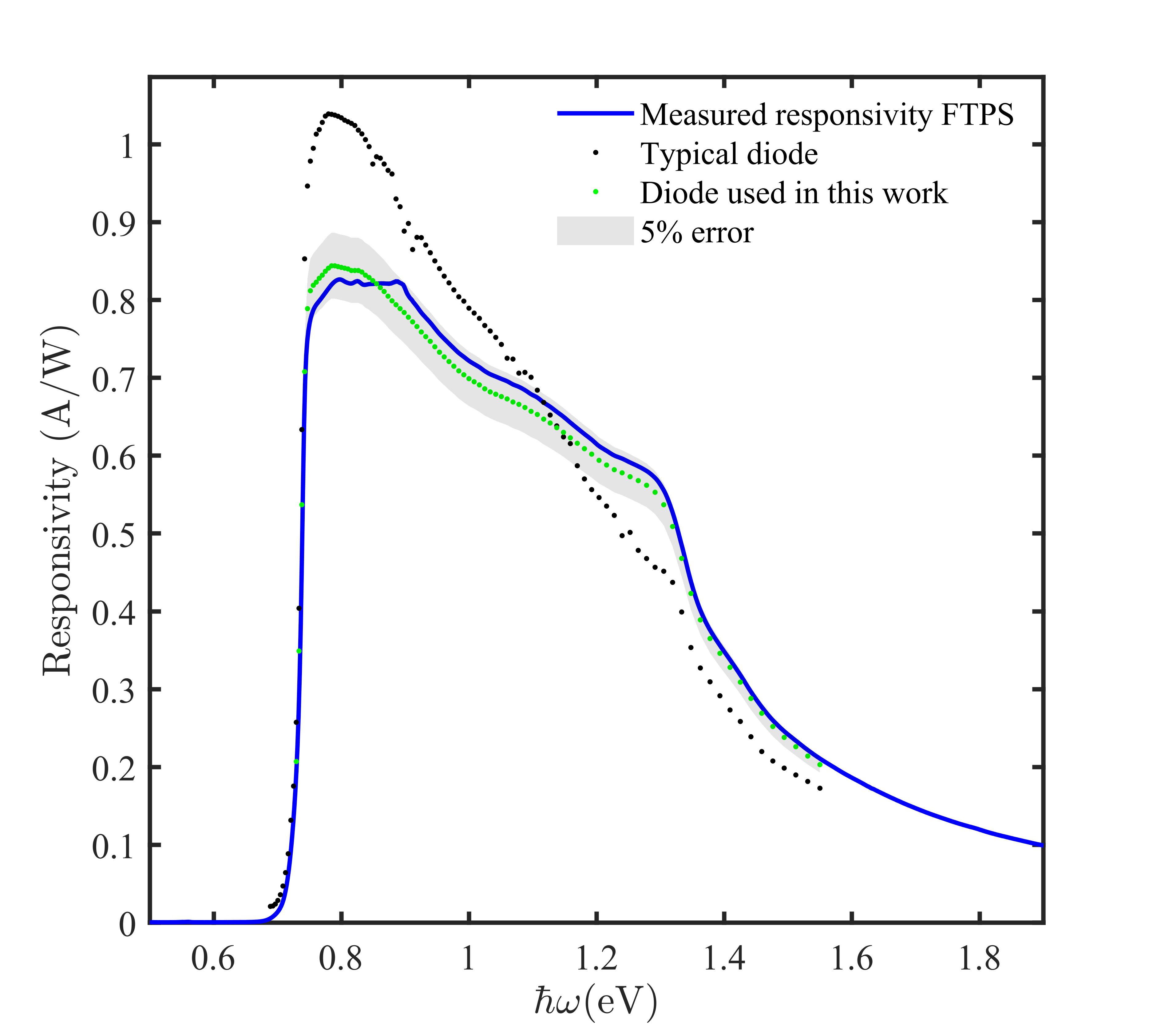}
\caption{Responsivity of InGaAs diode provided by Thorlabs for (black points) typical diode, (green points) our particular diode FGA21-CAL including 5\% error margin. }
\label{Fig3}
\end{figure} 
We resolved the effects of spectrally unequal white-light intensity by normalizing the photocurrent spectrum by the white-light power spectrum $P_{\mathrm{exc},j}$. We measured the white-light power $P_\mathrm{bp}$ behind the bandpass filter and determined the responsivity $R_\mathrm{bp}=I_\mathrm{tot,bp}/P_\mathrm{bp}$ at 1300~nm. We calibrated the responsivity using a bandpass filter (FB1300-12, Thorlabs) transmitting only wavelengths in the band (1300$\pm$12)~nm. The blue peak in Fig.~\ref{Fig1}~(a) depicts the bandpass FTIR spectrum of the InGaAs diode. Fig.~\ref{Fig1}~(b,c) shows the corresponding interferogram measured by the oscilloscope for two timescales. A detail in Fig.~\ref{Fig1}~(c) shows the interferogram maximum. It resembles the ideal interferogram for monochromatic light. The beating pattern on the timescale $\approx 100$~ms, Fig.~\ref{Fig1}~(b), is a fingerprint of the non-ideal monochromatic light. The envelope amplitude of the interferogram should not decay for coherent monochromatic light. The finite coherency leads to the amplitude decay for long delays of the two interfering beams in the Michelson interferometer. Because of the finite coherence time, the correct photovoltage is at zero delay time of the interfering beams. 

\begin{table}[b]
\caption{\label{Tab1} Quantities used to calculate the InGaAs diode's responsivity and photocurrent in FTPC spectra. }
\begin{ruledtabular}
\begin{tabular}{cccc}
Quantity&Value&Rel. Accuracy\\
\hline
Amplification\footnote{Trans-impedance preamplifier Newport 70710.}& (10$^4\pm10^2$) V/A& 1\% \\
Intensity\footnote{Tungsten white-light source, measured by the Thorlabs sensor S122C 700-1800~nm, 40~mW behind the Thorlabs bandpass filter FB1300-12.}&(6.27$\pm0.31$) $\mu$W &5\%  \\
Photovoltage\footnote{Measured using the Thorlabs, InGaAs FGA21-CAL, photovoltaic configuration, plugged in Bruker Vertex80v FTIR spectrometer}&(47$\pm 2$) mV& 4.3\%  \\
\hline
Responsivity&(0.75$\pm0.05$) A/W&6.7\% \\
Data-sheet Responsivity\footnote{Thorlabs, InGaAs FGA21-CAL}& (0.733$\pm$0.040) A/W& 5\%
\end{tabular}
\end{ruledtabular}
\end{table}
We summarize the results in Tab.~\ref{Tab1}. The responsivity of the InGaAs calibrated diode is within a 5\% error in agreement with the specification provided by the manufacturer (Thorlabs). We scale the responsivity $R_j$ in the whole measured spectral range by using the bandpass spectrum as a weighting function $w_i$, normalized such as $\sum_i w_i=1$, the index $i$ labels the experimental data on a discrete energy scale. The scaled FTPS responsivity spectrum $R_j$ is then
\begin{equation}
    R_j=\frac{u_j^{\mathrm{pwr}}}{\sum_i w_iu_i^{\mathrm{pwr}}}R_\mathrm{bp},
    \label{eq:responsivity}
\end{equation}
where $u_i^\mathrm{pwr}=u_i^\mathrm{exp}/P_{\mathrm{exc},i}$ is the spectral power $P_{\mathrm{exc},i}$ normalized FTPS signal $u_i^\mathrm{exp}$. We show the responsivity of the InGaAs diode in Fig.~\ref{Fig3}. The agreement with the specification is within the full spectral range at the level of 5\%. We note here that it is essential to use a calibrated photodiode. The photodiodes differ from piece to piece, and their responsivity can be quite distinct from the typical diode characteristics, as shown in Fig.~\ref{Fig3}. The major source of error is the absolute measurement of the optical power and measurement of the white light spectrum.

\begin{figure}[t!]
\centering
\includegraphics[width=8cm]{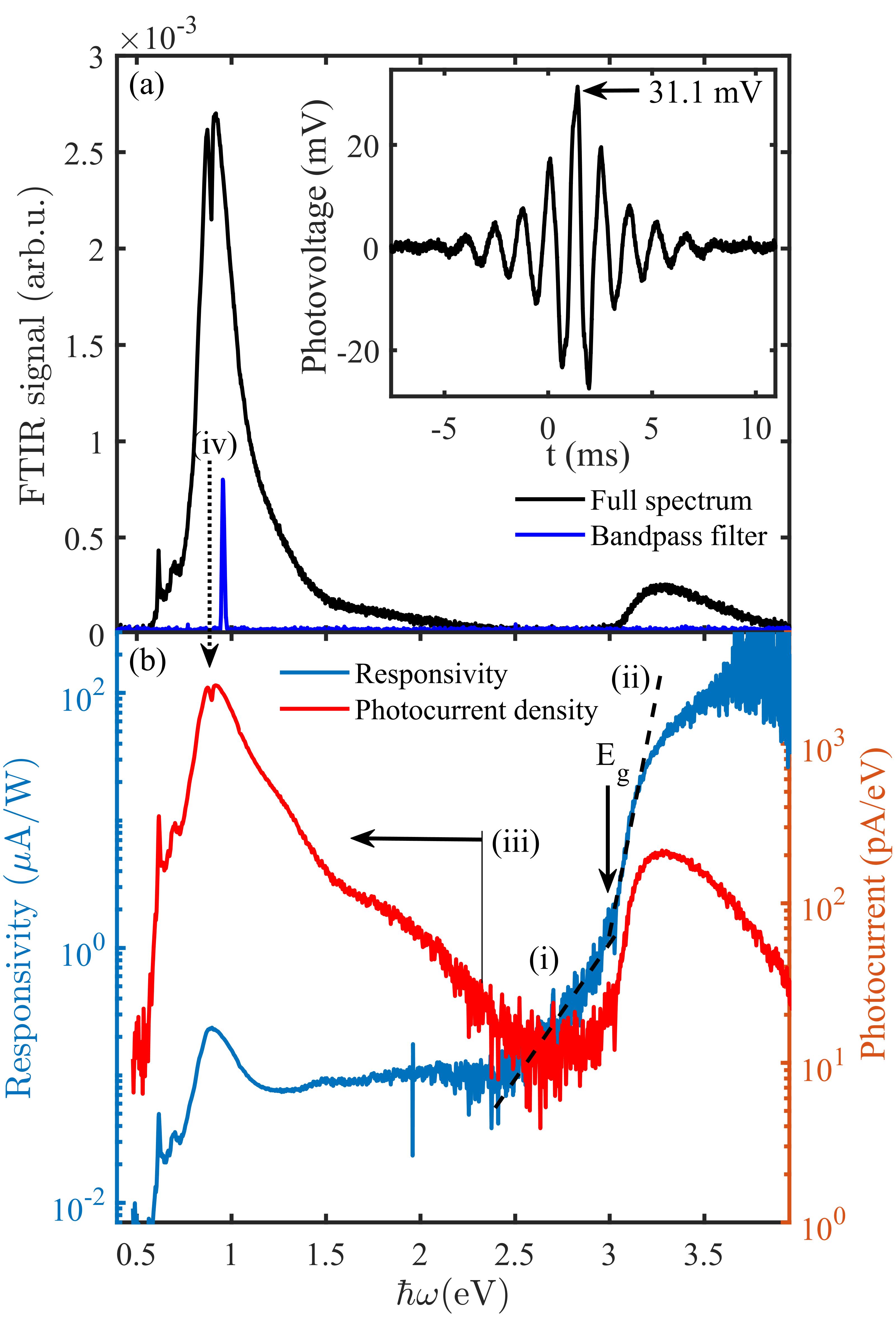}
\caption{(a) As-measured FTPC spectrum of the SiC interdigital sensor (black curve) without and (blue curve) with a bandpass filter. The inset shows the interferogram of the full spectrum without a bandpass filter. This data is the oscilloscope-measured interferogram behind the trans-impedance preamplifier. (b) Data scaled to (left scale, blue curve) responsivity and (right scale, red curve) photocurrent spectral density.}
\label{Fig4}
\end{figure}

Next, we show the responsivity scaling of weak signals below the sensitivity of standard oscilloscopes. We use the link between the integrated FTPC spectrum (calculated from the measured data - in arb.u.) and the total photovoltage ($U_\mathrm{ph, max}^{(i)}$). The ratio of integrated FTPC spectrum without bandpass filter $I_\mathrm{tot}$ and with bandpass filter $I_\mathrm{bp}$ has to equal the ratio of the integrated photovoltages measured with and without the filter. Hence, the expression
\begin{equation}
    \frac{I_\mathrm{tot}}{I_\mathrm{bp}}=\frac{V_\mathrm{tot}}{V_\mathrm{bp}}
    \label{eq:ratios}
\end{equation}
can be used to calculate the unknown $V_\mathrm{bp}$ at which we can reliably measure the optical power $P_\mathrm{bp}$. We verify Eq.~(\ref{eq:ratios}) using the sensitive InGaAs photodiode in Tab.~\ref{Tab:comparison}. Indeed, the ratio of integrated intensities equals the ratio of photovoltages within 5\% experimental error. Using the homodyne lock-in detection, we also verified Eq.~(\ref{eq:ratios}) for our low-responsivity SiC sample. We excite the photocurrent by the chopped white light exiting the FTIR spectrometer, and we measure the integral photovoltage $U_\mathrm{rms, back}=24.8(5)$~mV without and $U_\mathrm{rms, back}=0.29(5)$~mV with the bandpass filter. The ratio 85(15) fits the ratio of integrated FTPC signals 85.3, Tab.~(\ref{Tab:comparison}). We conclude that our assumption on Eq.~(\ref{eq:ratios}) is valid. The measured photoresponse of the 6H-SiC sample allows scaling the FTPC spectrum to photocurrent density. The measured optical power behind the bandpass filter $P_\mathrm{bp}=6.3(3)~\mu$W allows calculating the responsivity spectrum, Eq.~(\ref{eq:responsivity}). Fig.~\ref{Fig4}~(b) shows the photocurrent and responsivity of the 6H-SiC interdigital sensor.

The responsivity shows the onset of the optical absorption for photon energies $\hbar\omega$ larger than bandgap $E_\mathrm{g}=3.06$~eV. This absorption onset reflects the expected bandgap of the 6H-SiC polytype. These interband transitions are one to two orders of magnitude stronger than the impurity-band transitions in the responsivity spectrum. The indirect bandgap of SiC appears as two distinct slopes of the absorption edge, shown by dashed lines (i) and (ii) in Fig.~\ref{Fig4}~(b). The low-energy/high-energy part (i)/(ii) corresponds to the photon absorption assisted by the phonon's simultaneous absorption/emission. The transitions from the deep-level states to the conduction band dominate the absorption for energies $\hbar\omega<E_\mathrm{g}$. The Fermi level $E_\mathrm{F}\approx E_\mathrm{c}-0.9$~eV. The photoconductivity peak in the vicinity of 1~eV is related to the states pinning the Fermi level, thus compensating for the background doping in the semi-insulating SiC.

We point out here that FTPC spectroscopy is a sensitive technique providing valuable information about deep levels in semiconductors and interband transitions. The photocurrent sensitivity on the order of 100's~fA is related to the narrow bandwidth provided by the Fourier transform. The set spectral resolution and the scanner velocity give the Fourier transform bandwidth. The resolution 8~cm$^{-1}$ and scanner velocity 1650~Hz used in our experiments provide bandwidth below 1~Hz. Such a narrow bandwidth significantly limits noise level, allowing us to measure low photocurrent.

The primary issue of the scaling method described here is the case of weak signals. When the oscilloscope cannot measure the interferogram directly, Method 1 requires two assumptions. First, the beamsplitter in the Michelson interferometer splits the beam precisely in the ratio of 50:50. Second, the interfering wavelengths show zero delays at the same mirror position. The non-zero delay between the two beams might appear when the dispersion of the index of refraction of the beamsplitter is too large. The construction of the beamsplitter in the FTIR spectrometer Vertex80v compensates the dispersion by assuring that both interfering beams travel the same distance through the beamsplitter and experience the same number of transmissions and reflections under the same angles. However, the non-ideal construction can lead to slightly different optical paths. This nonideality leads to spatially shifted interferogram maxima for different wavelengths. In such a situation, the maximal photovoltage would be a meaningless quantity. However, we showed that for the NIR to VIS spectral range, using CaF$_\mathrm{2}$ beamsplitter, the proposed scaling leads to the errors within an order of magnitude (factor 2, see Tab.~(\ref{Tab:comparison}), $U_\mathrm{max}=31.1$~mV and $U_\mathrm{back,pk-pk}$=66.2(5)~mV).  

We described two methods to scale the FTPC spectra to photocurrent spectral density (A/eV). We also presented a method for responsivity (A/W) scaling. We verified the method on a NIST traceable calibrated InGaAs photodiode and applied the technique to 6H-SiC interdigital sensors. We also described an extension of the method to calibrate weak-signal spectra. The relation between the integrated FTPC signal and maximal interferogram voltage allows such scaling. 

We acknowledge financial support from the Czech Science Foundation under project 19-12052S. CzechNanoLab project LM2018110, funded by MEYS CR, is also gratefully acknowledged for the financial support of the sample fabrication at CEITEC Nano Research Infrastructure.

\bibliography{MainTextKunc}

\end{document}